\documentclass[11pt]{article}

\usepackage[margin=1.0in]{geometry}
\usepackage{authblk}
\usepackage{background}

\usepackage{amsmath}
\usepackage{amssymb}
\usepackage{bm}
\usepackage{color,soul}
\usepackage{hyperref}
\usepackage{siunitx}
\usepackage{subcaption}

\SetBgContents{Preprint}
\SetBgScale{1}
\SetBgAngle{0}
\SetBgOpacity{1}
\SetBgColor{black}
\SetBgPosition{current page.south west}
\SetBgHshift{1.0in}
\SetBgVshift{0.62in}

\begin{document}
\title{Entropy-based feature selection for capturing impacts in Earth system models with extreme forcing}

\author[1,*]{Jerry Watkins}
\author[2]{Luca Bertagna}
\author[2]{Graham Harper}
\author[2]{Andrew Steyer}
\author[1]{Irina Tezaur}
\author[2]{Diana Bull}

\affil[1]{\footnotesize Sandia National Laboratories, Livermore, CA, USA.}
\affil[2]{\footnotesize Sandia National Laboratories, Albuquerque, NM, USA.}
\affil[*]{\footnotesize Corresponding author; email: {\tt jwatkin@sandia.gov}.}

\begin{titlepage}
\maketitle
\thispagestyle{empty}
\begin{abstract}
This paper presents the development of a new entropy-based feature selection method for identifying and quantifying impacts. Here, impacts are defined as statistically significant differences in spatio-temporal fields when comparing datasets with and without an external forcing in Earth system models. Temporal feature selection is performed by first computing the cross-fuzzy entropy to quantify similarity of patterns between two datasets and then applying changepoint detection to identify regions of statistically constant entropy. The method is used to capture temperate north surface cooling from a 9-member simulation ensemble of the Mt. Pinatubo volcanic eruption, which injected 10 Tg of SO\textsubscript{2} into the stratosphere. The results estimate a mean difference decrease in near surface air temperature of $-0.560$K with a 99\% confidence interval between $-0.864$K and $-0.257$K between April and November of 1992, one year following the eruption. A sensitivity analysis with decreasing SO\textsubscript{2} injection revealed that the impact is statistically significant at 5 Tg but not at 3 Tg. Using identified features, a dependency graph model composed of 68 nodes and 229 edges directly connecting initial aerosol optical depth changes in the tropics to solar flux and temperature changes before the temperate north surface cooling is presented.
\end{abstract}
\end{titlepage}

\section{Introduction} \label{sec:intro}
Large external forcing to the Earth's climate system has the potential to trigger extreme events like heatwaves and droughts. For example, tipping events \cite{armstrong2022exceeding} can be modeled as large external forcing that can cause significant damage to local communities and infrastructure. It is crucial for Earth systems models to accurately capture and predict these regional impacts to design mitigation and adaptation strategies. Unfortunately, the cascading effects of an external forcing on the climate system are not well understood. Fundamental research is needed to develop methods that can attribute regional impacts to an external forcing.

Careful interrogation of climate data is required to develop a model that can uncover a source-to-impact pathway, a representation of the relationships and interactions between a set of spatio-temporal quantities of interest when the model is subjected to an external forcing. A data-driven approach seems well-suited to provide answers as long as there are enough data to train such a model. Unfortunately, generating climate simulation data can be expensive, and the data suffers from two common problems: high variability and high dimensionality. The nonlinear coupling of processes in the Earth system produce a complex signal when evaluating quantities of interest such as temperature and precipitation. Earth system model simulations can produce hundreds of spatio-temporally varying fields, all coupled together to produce a network of highly-correlated, nonlinear inter-dependencies. The nonlinear, chaotic nature of the data also causes many linear regression models to fail, exemplifying the need for models that capture nonlinear dynamics.

This article explores the development of an entropy-based feature selection method to reduce the complexity of constructing data-driven surrogate models for attribution. Feature selection is the process of selecting only the most important features for model development to improve accuracy, reduce training time, and construct a more interpretable model. In this context, it refers to a dimension reduction in space-time and a selection of the most relevant variables to explain the effects of a regional impact. 

\subsection{Previous related work} \label{subsec:intro-prev}
Feature selection is a widely researched topic in machine learning. There are a few relevant examples of its use with observational data in weather and climate. In \cite{feng2017data}, a feature selection process which included principal component analysis, Granger causality tests, autocorrelation analysis and recursive feature elimination improved wind forecast accuracy by up to 30\%. In \cite{li2022novel}, Li developed a feature selection technique based on conditional mutual information that performed better than correlation analysis and mutual information maximum for drought prediction models. In \cite{fang2022development}, Fang showed that performing Shapley additive global importance (SAGE) for regional feature selection in air pollution data resulted in a more interpretable, accurate model when compared with a model without feature selection. In contrast, this paper focuses on feature selection in Earth system model simulation data, which typically have a fixed domain in time and space.

Information entropy \cite{shannon1948mathematical} is a measure of uncertainty associated with a quantity within a probability distribution. Given a discrete time series, entropy can be approximated to measure the ``complexity" in the signal and has been used extensively in many fields, including weather and climate, to identify structure in nonlinear dynamical systems. For example, in \cite{jin2016applicability}, Jin showed that approximate entropy was able to detect abrupt changes in temperature and precipitation. Ikuyajolu used recurrence entropy to quantify predictability in climate models \cite{ikuyajolu2021information}. In \cite{karevan2018transductive}, a cluster-based sample entropy method was developed to perform feature selection in weather forecasting. In contrast, this article focuses on modeling the climate system with and without an extreme forcing where we can apply cross-entropy methods to paired datasets.

This article also explores the construction of directed graphs to explain climate impact pathways with verification on a Mt. Pinatubo eruption simulation ensemble. A similar approach is found in \cite{nichol2024space} where a grid-cell-level analyses is performed to learn local causal dynamical structure in space-time data. In \cite{Peterson:2024}, a weighted directed graph is constructed from pair-wise feature importance from random forest regression. For a comprehensive summary of other recently-developed data-driven methods, the reader is referred to \cite{CLDERA_SAND} and the references there-in.

\subsection{Main contributions} \label{subsec:intro-cont}
The high frequency data generated from Earth system model simulations of extreme events motivates the need for a feature selection method that can isolate source-to-impact pathways. In this article, a new, entropy-based feature selection method is introduced that attempts to focus on the most important features related to a pathway by constructing a dependency graph model. The novelty in the method is in its use of cross-fuzzy entropy \cite{xie2010cross} and changepoint detection to isolate regions of constant entropy between a forced and unforced dataset. The method is tested on a Mt. Pinatubo eruption simulation ensemble where the goal is to isolate spatio-temporal regions of high importance given a source and impact. A sensitivity analysis with respect to SO\textsubscript{2} magnitude is also performed leading to new insights in the capability of Earth system models in capturing impacts from smaller stratospheric injections. The methods introduced focus on capturing the cascading effects of an external forcing in climate models but are extensible to other nonlinear dynamical systems with high frequency spatial-temporal data.

The remainder of this article is organized as follows. Section~\ref{sec:impact} provides a detailed overview of the feature selection method including how the input and output data is generated. Section~\ref{sec:results} gives a numerical example of how the method can be used to select important features in climate data using a Mt. Pinatubo eruption simulation ensemble. Lastly, Section~\ref{sec:conc} offers concluding remarks on how the method can be further improved and utilized.

\section{Entropy-based feature selection} \label{sec:impact}
Feature selection seeks to identify the most important features in a data set for model development. In the context of high frequency Earth system model data, there are two main objectives:
\begin{enumerate}
\item improve the ability to detect quantifiable and explainable features in the presence of high variability; and
\item reduce the degrees of freedom needed to construct surrogate models for attribution.
\end{enumerate}
The data contains hundreds of spatio-temporal fields with nonlinear dependencies. In order to identify the most important features of an external forcing, the following workflow is proposed:
\begin{itemize}
\item \textbf{Data generation:} generate an Earth system model simulation ensemble with and without forcing. This allows for the creation of a model that can look for differences between the two datasets and identify key features caused by the forcing. 
\item \textbf{Regional labels:} perform dimension reduction in space using scientific regions of interest. Most scientific studies focus on key regions to explain impacts to local communities. For example, climate zonal bands or regions based on the sixth Intergovernmental Panel on Climate Change (IPCC) assessment report (AR) \cite{iturbide2020update}.
\item \textbf{Temporal feature selection:} perform dimension reduction in time using intervals of constant cross-entropy. This work focuses on isolating time intervals unique to the external forcing.
\item \textbf{Impact statistics:} perform statistical analysis to quantify difference between the forced and unforced model. Quantifying the difference helps identify and rank abnormal features in the dataset.
\item \textbf{Pathway graph model:} construct a dependency graph based on known dependencies. There are fundamental physical processes inherent in the model data that can help constrain the pathway graph model.
\end{itemize}
The implementation of each process within the context of our Mt. Pinatubo exemplar is described in more detail below.

\subsection{Data generation} \label{subsec:impact-data}
A detailed explanation of the limited variability ensemble used in this study is provided in \cite{ehrmann2024identifying}. The following is a short summary of the data generated.

The volcanic eruption of Mt. Pinatubo occurred on June 15th, 1991 at the latitudinal and longitudinal coordinates (\ang{15;08;30}N, \ang{120;21;00}E) and injected around 20 Tg of sulfate into the atmosphere with around 10 Tg of SO\textsubscript{2} remaining in the stratosphere \cite{Guo2004,Kremser2016}. This event was simulated using the Energy Exascale Earth System Model (E3SM) version 2 \cite{golaz2022doe} with prognostic aerosol modifications\footnote{\url{https://github.com/sandialabs/cldera-e3sm}} \cite{brown2024validating} and fully coupled atmosphere, land, ocean and sea-ice processes. The model was initialized on June 1st, 1991 based on observed climate conditions and emission of 10 Tg of SO\textsubscript{2} at an altitude between 18 and 20 km on June 15th, 1991 according to the procedure detailed in \cite{brown2024validating}. A limited variability ensemble in which select natural modes matched historical conditions at the time of eruption (both the El Ni\~{n}o Southern Oscillation and the Quasi-biennial Oscillation \cite{ehrmann2024identifying}) is constructed by perturbing the initial temperature field using a reproducible random number generator. Each ensemble member was paired with a counterfactual simulation where the Mt. Pinatubo eruption was removed. The paired ensemble members are exactly the same until the eruption at which point they start to diverge from one another due to the chaotic nature of the atmosphere. The simulations were run up to January 1st, 1999 on an approximately 1 degree resolution ne30 grid where ne30 refers to the number of elements along each edge of a cubed-sphere grid. Regional means were computed in situ at 30 minute time intervals using {\tt CLDERA-Tools}\footnote{\url{https://github.com/sandialabs/cldera-tools}}, a library written in C++ that supports E3SM field extraction to compute quantities of interest while the simulation is running \cite{steyer2024situ}.

As in \cite{hart2024stratospheric} the emissions detailed in \cite{brown2024validating} were scaled for each ensemble member to create a pattern of responses sensitive to the degree of forcing ~\cite{Arnell2019,Osborn2016,Herger2015}. The emissions were scaled by 0, 0.1, 0.3, 0.5, 0.7, 1, 1.3 and 1.5 to achieve a range of SO\textsubscript{2} magnitudes from 0 Tg (the ``counterfactual'' or hereafter CF) to 15 Tg ($\sim$ 50\% greater than the estimated historical eruption of 10 Tg of stratospheric SO\textsubscript{2}). Other than the magnitude scaling, all other aspects of the eruption (date, location, and injection height) remain the same.

\subsection{Regional labels} \label{subsec:impact-regions}
The data are reduced in space by computing labeled regional means in {\tt CLDERA-Tools} \cite{steyer2024situ}. First, a masking file is constructed using a {\tt NetCDF} file containing the latitudinal and longitudinal coordinates of the ne30 grid. The masking file contains a list of indices with a dimension equal to the total number of columns in the grid where each index is associated with a non-overlapping region.

\begin{figure}[htbp]
\centering
\includegraphics[width=0.85\textwidth]{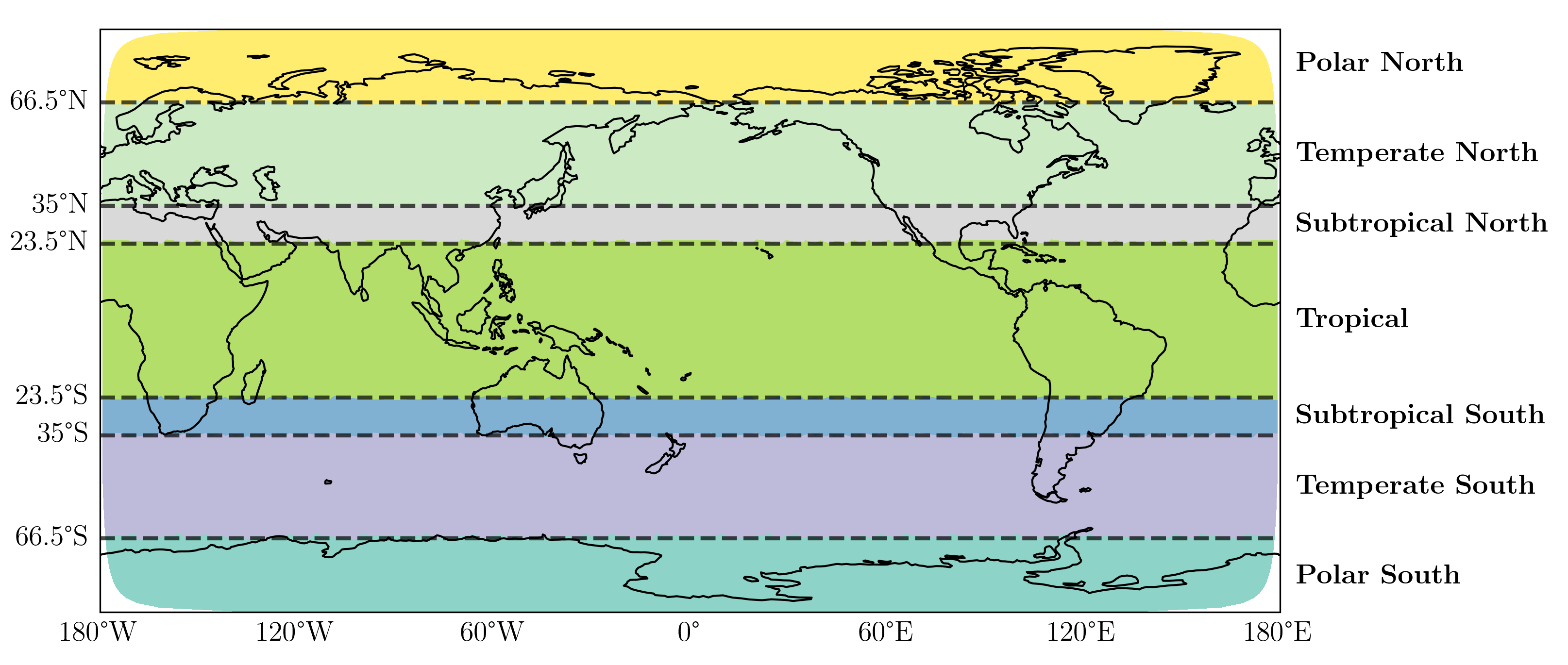}
\caption{Zonal regions are masked employing the atmospheric Hadley and Ferrel cells as the main distinguisher of regions, and means are computed and labeled on a 1 degree resolution grid. The Mt. Pinatubo eruption occurs in the tropical band around $15^{\circ}$N and $60^{\circ}$W on this grid.}
\label{fig:impact-zonal}
\end{figure}

{\tt CLDERA-Tools} computes regional means using multiple masking files. Global, zonal, and AR6 IPCC \cite{iturbide2020update} means are computed for 61 fields. Means over zonal regions, shown in Figure \ref{fig:impact-zonal},  are used in this study.

\subsection{Temporal feature selection} \label{subsec:impact-time}
Given two time series, one with the eruption and one without (the counterfactual), the objective is to identify discrete points in time where the statistical similarity of patterns between the two signals significantly diverges in order to capture time intervals where an external forcing may have caused a significant change. The first step is to apply cross-fuzzy entropy \cite{xie2010cross}, which measures the degree of pattern synchrony. Let,
\begin{equation}
\begin{alignedat}{1}
u &= \left\{u_1,  u_2, \ldots, u_N\right\}, \\
v &= \left\{v_1,  v_2, \ldots, v_N\right\},
\end{alignedat}
\end{equation}
be spatial means of any field variable in the climate data where $u$ is a time series in the eruption dataset, $v$ is a time series in the counterfactual dataset, and $N$ is the total number of points in time. In order to capture higher frequency entropy shifts, the data are split into $M$ sliding windows of size $n$ and lag $p$,
\begin{equation}
\begin{alignedat}{2}
U_i &= \left\{u_{(i-1)p+1},  u_{(i-1)p+2}, \ldots, u_{(i-1)p+n}\right\}, \qquad &&i = 1,2,\ldots,M, \\
V_i &= \left\{v_{(i-1)p+1},  v_{(i-1)p+2}, \ldots, v_{(i-1)p+n}\right\}, \qquad &&i = 1,2,\ldots,M.
\end{alignedat}
\end{equation}

Focusing on computing entropy on the first window, $U_1 = \left\{u_1,  u_2, \ldots, u_n\right\}$ and $V_1 = \left\{v_1,  v_2, \ldots, v_n\right\}$, for simplicity, cross-fuzzy entropy is computed by first constructing the following set of vectors,
\begin{equation}
\begin{alignedat}{2}
X_i^m &= \left\{u_i,  u_{i+1}, \ldots, u_{i+m-1}\right\} - \bar{u}_i^m, \qquad &&i = 1,2,\ldots,n-m, \\
Y_j^m &= \left\{v_j,  v_{j+1}, \ldots, v_{j+m-1}\right\} - \bar{v}_j^m, \qquad &&j = 1,2,\ldots,n-m,
\end{alignedat}
\end{equation}
where $m$ is the embedding dimension that sets the size of the vector for comparison, and,
\begin{equation}
\begin{alignedat}{2}
\bar{u}_i^m &= \frac{1}{m}\sum_{l=0}^{m-1}u_{i+l}, \qquad &&i = 1,2,\ldots,n-m, \\
\bar{v}_j^m &= \frac{1}{m}\sum_{l=0}^{m-1}v_{j+l}, \qquad &&j = 1,2,\ldots,n-m,
\end{alignedat}
\end{equation}
are the baseline signals. The maximum absolute difference is used as the distance function for the two vectors, $X_i^m$ and $Y_j^m$,
\begin{equation}
d_{i,j}^m = d(X_i^m, Y_j^m) = \max_{k \in (0,m-1)}\left|u_{i+k} - \bar{u}_i^m - \left(v_{j+k} - \bar{v}_j^m\right)\right|,
\end{equation}
and an exponential function is used as the fuzzy function,
\begin{equation}
D_{i,j}^m = \mu(d_{i,j}^m) = \exp\left(-\frac{\left(d_{i,j}^m\right)^{r_2}}{r_1}\right),
\end{equation}
where $r = \{r_1, r_2\}$ are parameters of the fuzzy function. The two values,
\begin{equation}
\begin{alignedat}{1}
\phi^m &= \frac{1}{N-m}\sum_{i=1}^{N-m}\left(\frac{1}{N-m}\sum_{j=1}^{N-m}D_{i,j}^m\right), \\
\phi^{m+1} &= \frac{1}{N-m}\sum_{i=1}^{N-m}\left(\frac{1}{N-m}\sum_{j=1}^{N-m}D_{i,j}^{m+1}\right),
\end{alignedat}
\end{equation}
are then constructed to compute the cross-fuzzy entropy as,
\begin{equation}
s_1 = s(U_1,V_1) = \ln\phi^m - \ln\phi^{m+1}.
\end{equation}
This is computed for all windows using the {\tt EntropyHub} code \cite{flood2021entropyhub} so that,
\begin{equation}\label{eq:entropy}
s = \left\{s_1,  s_2, \ldots, s_M\right\}.
\end{equation}

The entropy time series provides a measure of pattern synchrony but does not identify sudden changes. For that, changepoint detection is used as developed in \cite{watkins2023performance} to capture regions of constant entropy. Changepoint detection performs hypothesis testing in sequential subsets of data to determine if a change has occurred. A subset of entropy is defined as,
\begin{equation}
s_{i:j} = \left\{s_i,  s_{i+1}, \ldots, s_j\right\},
\end{equation}
where $i$ and $j$ are lower and upper limits within the time series in Equation~\eqref{eq:entropy}. Hypotheses are constructed as,
\begin{equation}
\begin{alignedat}{2}
H_0 &: f_{1:\hat{\nu}-1} = f_{\hat{\nu}:M}, \qquad \forall &&\hat{\nu} \in \mathcal{K}, \\
H_A &: f_{1:\hat{\nu}-1} \neq f_{\hat{\nu}:M}, \qquad &&\hat{\nu} \in \mathcal{K},
\end{alignedat}
\end{equation}
where $f_{i:j}$ is the probability density function $\forall \hat{s} \in s_{i:j}$ and $\mathcal{K} = \left\{2, 3, \ldots, M\right\}$. This can be viewed as a generalized likelihood ratio test where $H_0$ states that all $\hat{s} \in s$  belong to a single probability distribution while $H_A$ states that there exists some $\hat{\nu}$ such that all $\hat{s} \in s_{1:\hat{\nu}-1}$ and $\hat{s} \in s_{\hat{\nu}:M}$ belong to two separate probability distributions, respectively. 

A two-sample $t$-test of $s_{1:\hat{\nu}-1}$ and $s_{\hat{\nu}:M}$ is performed to determine whether $\hat{\nu}$ is a potential changepoint. In order to perform multiple hypothesis tests, the Bonferroni correction \cite{bonferroni1936teoria} is used to adjust the significance level by $\alpha / K$, where $\alpha$ is the desired significance level and $K \leq M-1$ is the number of tests. Only the largest changes in the time series are chosen for testing. Multiple changepoints are found in the time series by applying a sequential algorithm, as described in \cite{watkins2023performance}. The result is a list of changepoints in the dataset, $s$, where the largest changes of entropy are found. This is then mapped back to the original time series using the midpoints of each entropy window so that,
\begin{equation}
\nu = \left\{\nu_1, \nu_2, \ldots, \nu_P\right\},
\end{equation}
are defined on the original datasets, $u$ and $v$, where $P$ is the total number of changepoints. This produces a list of subsets of $u$ and $v$ that correspond to a zeroth order approximation of cross-fuzzy entropy,
\begin{equation}
\begin{alignedat}{2}
u_{i:j} &= \left\{u_i,  u_{i+1}, \ldots, u_j\right\}, \qquad &&\{i,j\} = \{\nu_1,\nu_2\},\{\nu_2,\nu_3\},\ldots,\{\nu_{P-1},\nu_P\}, \\
v_{i:j} &= \left\{v_i,  v_{i+1}, \ldots, v_j\right\}, \qquad &&\{i,j\} = \{\nu_1,\nu_2\},\{\nu_2,\nu_3\},\ldots,\{\nu_{P-1},\nu_P\},
\end{alignedat}
\end{equation}
or $P-1$ features in the datasets, $u$ and $v$, that correspond to time intervals of constant entropy or similarity of patterns between the two datasets.

\subsection{Impact statistics} \label{subsec:impact-stats}
The Earth system model simulation ensemble used in this study is small so a $t$-statistic is used to compute uncertainty bounds and $t$-scores for impacts in the data. Here, an impact is defined as a statistically significant change to a spatial-temporally varying field when comparing datasets with and without an external forcing.

Given two subsets, $u_{e,i:j}$ and $v_{e,i:j}$, where $e = 1,2,\ldots,E$ and $E$ is the ensemble size, the mean difference over time is computed for all ensemble members and features as,
\begin{equation}
w_{e,\{i,j\}} = \tfrac{1}{(j-i+1)}\sum_{l=i}^{j}\left(u_{e,i:j} - v_{e,i:j}\right), \qquad \{i,j\} = \{\nu_1,\nu_2\},\{\nu_2,\nu_3\},\ldots,\{\nu_{P-1},\nu_P\},
\end{equation}
and the mean and standard error of the temporal mean difference over the ensemble is computed as,
\begin{equation}
\begin{alignedat}{2}
&\bar{w}_{\{i,j\}} = \tfrac{1}{E}\sum_{e=1}^{E} w_{e,\{i,j\}}, \qquad &&\{i,j\} = \{\nu_1,\nu_2\},\{\nu_2,\nu_3\},\ldots,\{\nu_{P-1},\nu_P\}, \\
&\text{SE}_{\bar{w}_{\{i,j\}}} = \frac{\text{SD}_{\bar{w}_{\{i,j\}}}}{\sqrt{E}}, \qquad &&\{i,j\} = \{\nu_1,\nu_2\},\{\nu_2,\nu_3\},\ldots,\{\nu_{P-1},\nu_P\},
\end{alignedat}
\end{equation}
where $\text{SD}$ is the standard deviation. The 99\% confidence interval (CI) for the mean difference is computed using a $t$-distribution,
\begin{equation}
\text{CI}_{\{i,j\}} = \bar{w}_{\{i,j\}} \pm t_{0.005} \text{SE}_{\bar{w}_{\{i,j\}}}, \qquad \{i,j\} = \{\nu_1,\nu_2\},\{\nu_2,\nu_3\},\ldots,\{\nu_{P-1},\nu_P\},
\end{equation}
where $t_{0.005}$ is the $t$-statistic. This corresponds to an impact estimate for each time interval with uncertainty bounds. The $t$-score for each feature is also computed,
\begin{equation}
\text{Score}_{\{i,j\}} = \frac{\bar{w}_{\{i,j\}}}{\text{SE}_{\bar{w}_{\{i,j\}}}}, \qquad \{i,j\} = \{\nu_1,\nu_2\},\{\nu_2,\nu_3\},\ldots,\{\nu_{P-1},\nu_P\}.
\end{equation}
This is used to determine the statistical significance of the impact.

\subsection{Pathway graph model} \label{subsec:impact-DAG}
The pathway from source-to-impact in Earth system model data is modeled using a directed acyclic graph (DAG). This follows the work from \cite{steyer2024situ}, where a time-dependent pathway DAG was used to represent source-to-impact pathways in an idealized volcanic eruption case referred to as HSW-V \cite{hollowed2024hsw}. In this context, a pathway is defined as a set of variables in space-time that interact to form a direct link between an external forcing and an impact.

A DAG is defined by its nodes and directed edges, $\mathcal{G} = \{\mathcal{Q}, \mathcal{E}\}$, where $\mathcal{Q} = \{q_1, q_2, \ldots \}$ is the set of all nodes and $\mathcal{E} = \{e_1, e_2, \ldots \}$ is the set of all edges. A pathway DAG is constructed by first defining nodes as impacts or features that meet a threshold for $t$-score,
\begin{equation}
q_{i:j} := \text{Score}_{\{i,j\}} > \epsilon, \qquad \{i,j\} = \{\nu_1,\nu_2\},\{\nu_2,\nu_3\},\ldots,\{\nu_{P-1},\nu_P\},
\end{equation}
where $\epsilon$ is the tolerance for statistical significance. Edges are created 
through the following set of physical constraints inherent in the model:
\begin{enumerate}
\item \textbf{Temporal constraint:} impacts propagate forward in time and can overlap; 
\item \textbf{Spatial constraint:} impacts propagate through neighboring regions; 
\item \textbf{Variable constraint:} impacts propagate through known or hypothesized variable dependencies.
\end{enumerate}
The edges define a set of plausible causal dependencies between nodes. Given these constraints, a ``full pathway DAG'', $\mathcal{G}_F$, can be constructed.

There are two additional traits of the dataset that can be used to further constrain the model: information about the source node where the external forcing was applied, and information about the final impact which is defined by the scientific study of interest. The ``impact pathway DAG'', $\mathcal{G}_I$, is constructed by adding the final impact constraint and including all upstream dependencies. Then, $\mathcal{G}_I$ can be further constrained by using a greedy best-first search algorithm to choose impact nodes with the largest $t$-score until the source node is found. This is referred to as the ``source-impact pathway DAG'', $\mathcal{G}_S$. This DAG helps identify impacts that have strong statistical significance and could potentially be large contributors to the final impact node.

Enforcing these spatio-temporal and variable constraints requires high frequency temporal data in order to ensure connections between nodes are generated. If the temporal frequency of field variables is too low, impacts may propagate further without the construction of edges. Since saving to disk all model quantities of interest at the same frequency as the model timestep may be prohibitive, in-situ techniques can be used to perform part of the calculations at runtime and reduce the volume of data that needs to be stored for post-processing analysis.
\section{Numerical Results} \label{sec:results}
The primary impact from the Mt. Pinatubo volcanic eruption was the change to radiative effects, which resulted in global mean surface cooling. Irregular regions of surface cooling were found in the tropical and mid-latitudes in the limited variability ensemble within the first year after the eruption \cite{ehrmann2024identifying}. In this section, the entropy feature selection method described in Section~\ref{sec:impact} is applied to the dataset to further analyze temperate north surface cooling and its association with downstream impacts from the Mt. Pinatubo eruption.

The data and parameters for the method were chosen based on an initial exploratory analysis on a separate five member, limited variability ensemble. The dataset includes daily, regional means for three variables as defined in Table~\ref{tab:res-variables}.
These variables define the so-called ``surface cooling pathway''. At the Earth's surface, we expect to see an increase in aerosol optical depth (AEROD\_v) due to the presence of sulfates, a  decrease in the amount of shortwave radiation reaching the surface (FSDSC), and, finally, a lowering of the temperature at the surface (TREFHT).
\begin{table*}[htbp]
\centering
\caption{Daily, regional mean variables used to identify Mt. Pinatubo as the source of regional surface cooling impacts.}
\label{tab:res-variables}
\begin{tabular}{p{4.0cm}p{8.0cm}}
\hline\noalign{\smallskip}
E3SM variable name & Description \\
\noalign{\smallskip}\hline\noalign{\smallskip}
AEROD\_v & Total aerosol optical depth in the visible band \\
FSDSC & Clearsky downwelling solar flux at the surface \\
TREFHT & Near surface air temperature \\
\noalign{\smallskip}\hline\noalign{\smallskip}
\end{tabular}
\end{table*}

For temporal feature selection, a window size of $n = 30$ with a lag size of $p = 9$ was chosen to capture seasonal transitions while also obtaining enough data points for accurate changepoints. For cross-fuzzy entropy, an embedding dimension of $m=2$ was used and the default parameters, $r = \{0.2, 2\}$, were used for the fuzzy function. The changepoint detection algorithm use the same parameters as used in \cite{watkins2023performance}. In all cases that follow, entropy is computed by using 9-member ensemble means for the eruption and counterfactual datasets, respectively.

\subsection{Temperate north surface cooling} \label{subsec:results-temp}
\begin{figure}[htbp]
\centering
\includegraphics[width=0.8\textwidth]{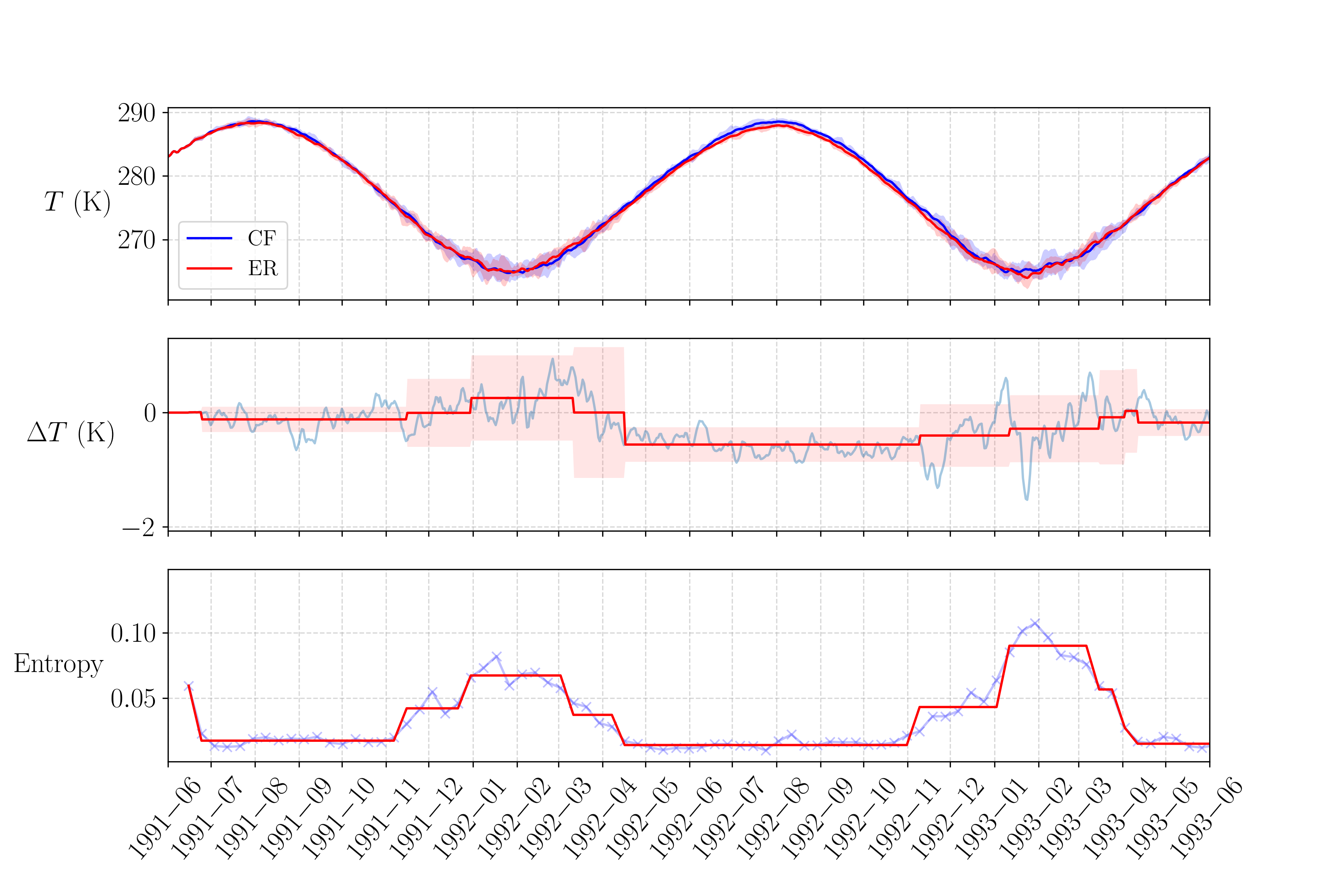}
\caption{The daily mean near surface air temperature in the temperate north, TREFHT, is computed in a 9-member ensemble with and without the Mt. Pinatubo eruption. The top figure shows the mean temperature across the ensemble for the eruption (ER) and counterfactual (CF) datasets with the minimum and maximum values shaded. The middle figure shows the mean difference of the two sets in blue, the temporal mean in each entropy-based time interval in red, and a 99\% confidence interval (CI) in a red shade. Lastly, the bottom figure shows the entropy in blue and the time intervals of constant entropy from changepoint detection in red. A statistically significant impact is found between April and November of 1992 and was computed as $-0.560$K with a 99\% CI between $-0.864$K and $-0.257$K.}
\label{fig:res-TREFHT}
\end{figure}

The near surface air temperature signal, TREFHT, in the temperate north suffers from a high amount of variability in the simulation ensemble. This serves as a good numerical test for the entropy-based method to determine its ability to capture impacts in the presence of noise. Figure~\ref{fig:res-TREFHT} shows two years of the TREFHT signal in the eruption and counterfactual ensemble along with the mean difference and the entropy. 
A total of $10$ statistically significant changepoints were found in the entropy signal in the bottom panel of Figure~\ref{fig:res-TREFHT}. Despite the high amount of variability, the method is able to track seasonal shifts in irregularity of the signal by grouping longer time intervals in the summer months and shorter time intervals in the winter months. In this case, it is not clear whether a shift in similarity of patterns between the two time series is caused by the eruption but statistically significant changes in TREFHT due to the eruption were found between April 17, 1992 and November 10, 1992. The mean difference in temperature was computed as $-0.560$ K with a 99\% confidence interval between $-0.864$ K and $-0.257$ K.

\begin{figure}[htbp]
\centering
\includegraphics[width=0.8\textwidth]{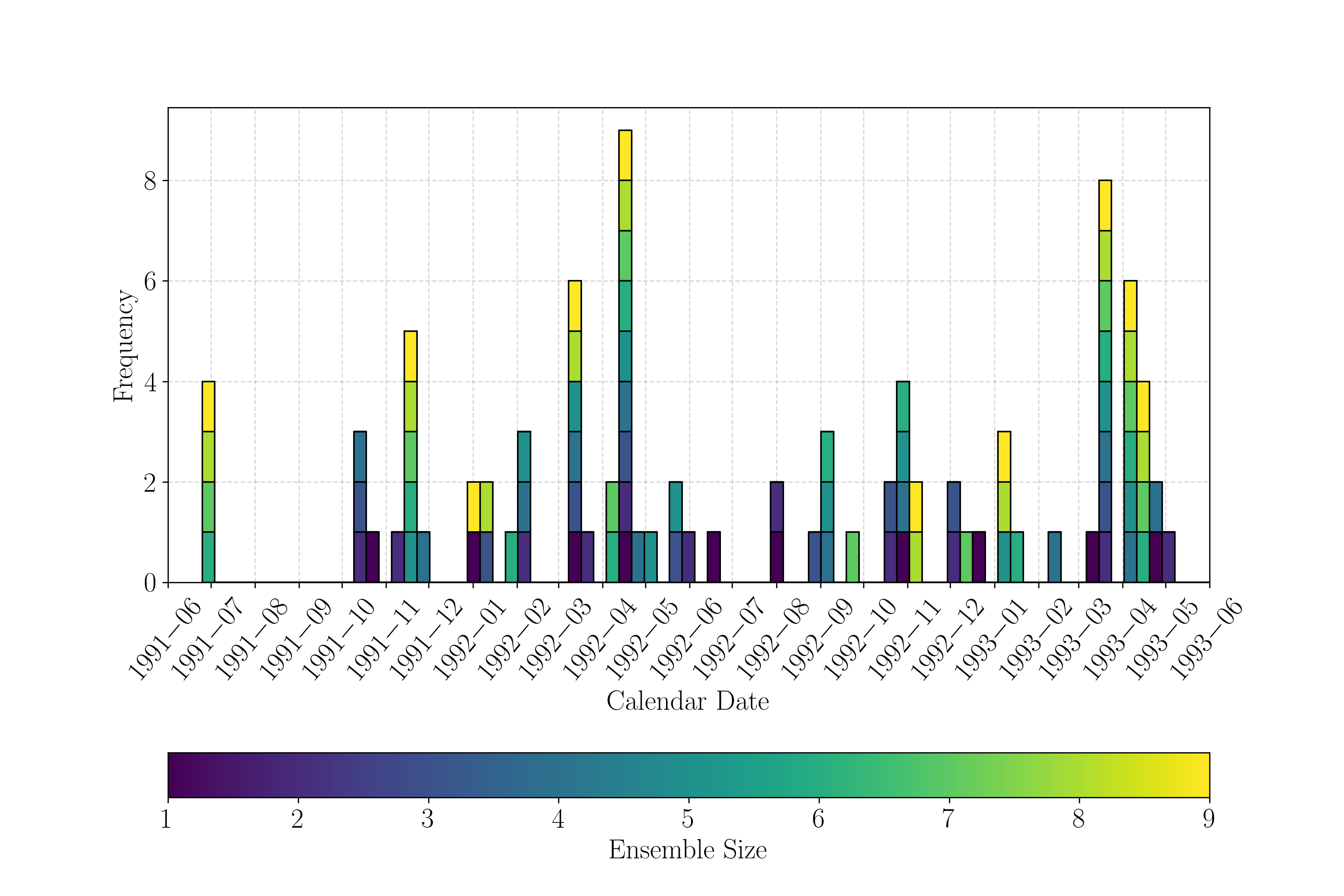}
\caption{The daily mean near surface air temperature in the temperate north, TREFHT, is computed with and without the Mt. Pinatubo eruption using ensemble sizes of $E = 1,2,\ldots,9$. At each ensemble size, entropy changepoints are computed and added to the histogram using eruption and counterfactual ensemble means. Here, the frequency is defined as the number of times the changepoint was found across all ensemble sizes. Different colors are used to show the changepoints found at each ensemble size. The $10$ changepoints found in the $9$-member ensemble in Figure~\ref{fig:res-TREFHT} were visited at least twice in this case, with the estimated start date of the largest impact in April stable across all ensemble sizes.}
\label{fig:res-chgpts}
\end{figure}

Another property to consider is the robustness of the entropy changepoints as more ensemble members are introduced to improve accuracy. In the bottom panel of Figure~\ref{fig:res-TREFHT}, the entropy changepoints are computed from 9 ensemble members, Figure~\ref{fig:res-chgpts} explores how these entropy changepoints would change if there where less ensemble members.
Here, entropy changepoints are computed multiple times using eruption and counterfactual ensemble means of ensemble sizes of $E = 1,2,\ldots,9$. A changepoint is considered more stable if the same changepoint is found across all ensemble sizes. In this case, the $10$ changepoints found in the $9$-member ensemble in Figure~\ref{fig:res-TREFHT} were visited at least twice. In particular, the April 17, 1992 changepoint, where the beginning of the impact was estimated, did not deviate as ensemble size was increased. The estimated end date of the impact, November 10, 1992, was less certain with a distribution that favored an earlier end date. Note that all estimated dates have a 9 day window of uncertainty due to the 9 day lag used to compute the entropy.

\begin{figure}[htbp]
\centering
\includegraphics[width=0.8\textwidth]{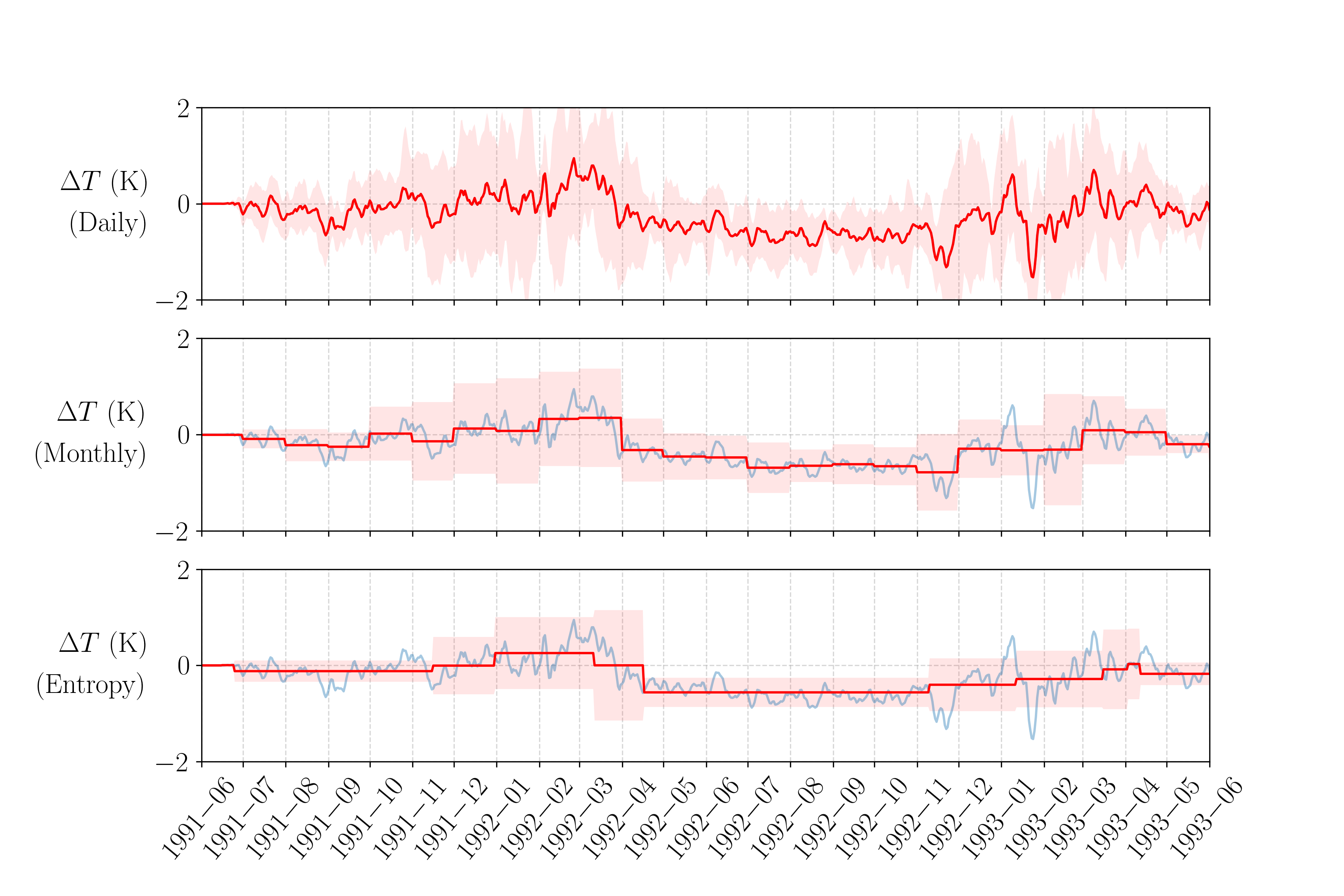}
\caption{The mean difference of near surface air temperature in the temperate north, TREFHT, is computed in a 9-member ensemble with and without the Mt. Pinatubo eruption. The daily, monthly and entropy-based temporal means are shown in red while a 99\% confidence interval is shown as a red shade. When the shade does not cover the zero axis, it is considered a statistically significant change associated with the eruption. There is a higher likelihood of false positives in the daily and monthly means.}
\label{fig:res-time-varying}
\end{figure}

TREFHT impacts can also be captured by computing statistics on daily and monthly means, but the high frequency of the data can often leads to a multiplicity problem where false positives are more likely to occur. Figure~\ref{fig:res-time-varying} shows the mean difference of the signal overlaid with a 99\% confidence interval using a daily, monthly and entropy-based temporal mean.
When the overlay does not cover the zero axis, it is considered a statistically significant change associated with the eruption. The daily means find many such points which may be associated with a false positive. Also, since many of them are not persistent, they are more likely to be associated with smaller impacts, which are not as relevant to other impacts in a pathway. The monthly means capture a similar statistically significant cooling when compared to the entropy means but the features are split by month which leads to more tests and a higher likelihood of false positives. Also, the estimated start date of the impact is significantly later in the monthly data when compared to the entropy-based means leading to less data that can be associated to the eruption.

\begin{figure}[htbp]
\centering
\includegraphics[width=0.8\textwidth]{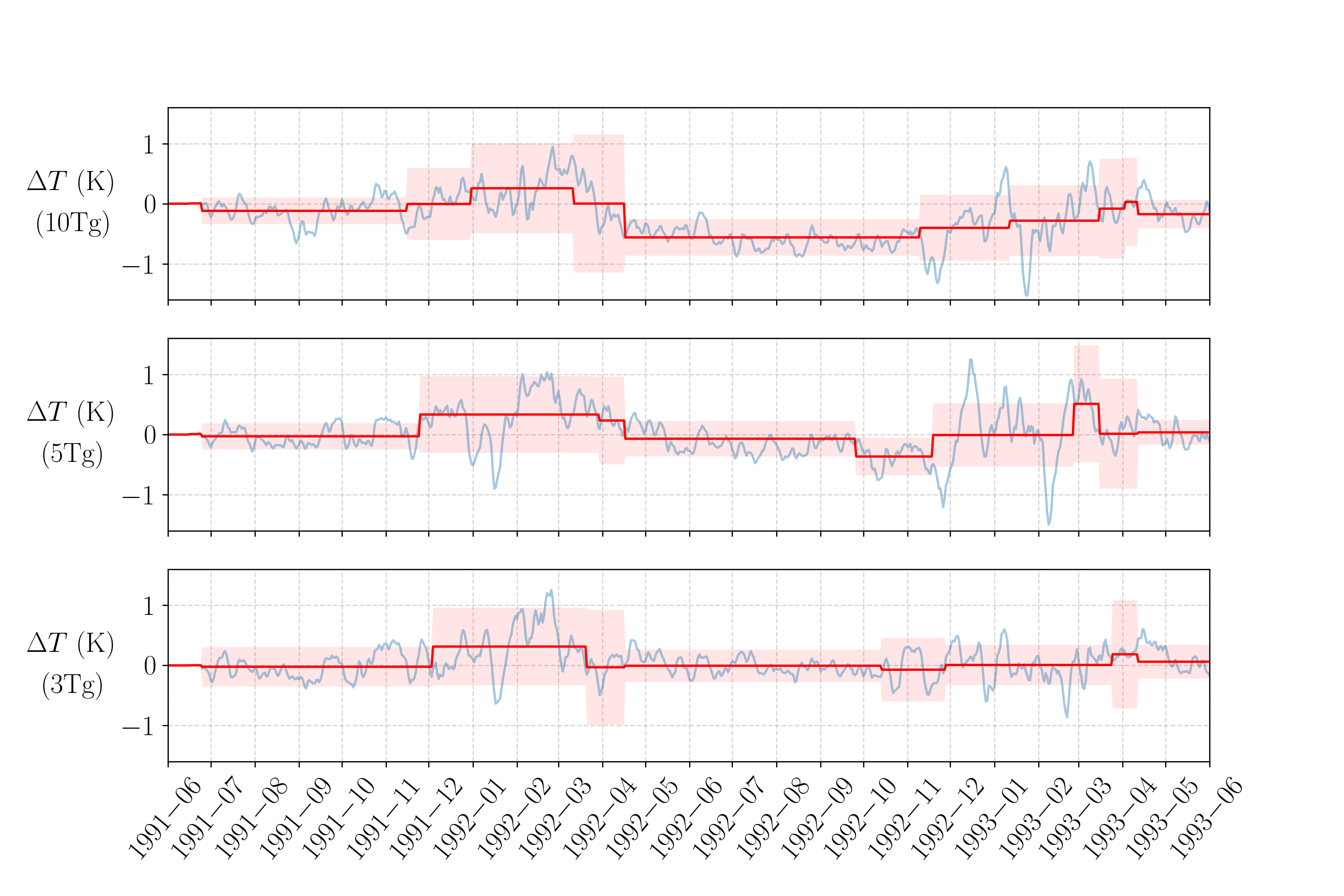}
\caption{The daily mean difference of near surface air temperature in the temperate north, TREFHT, is computed in a 9-member ensemble with and without eruptions of varying magnitude. The entropy-based temporal mean is shown in red while a 99\% confidence interval is shown as a red shade in a 10, 5 and 3 Tg mass eruption. When the shade does not cover the zero axis, it is considered a statistically significant change associated with the eruption. There is a small statistically significant cooling between September and November of 1992 in the 5 Tg eruption but no impact was found in the 3 Tg eruption.}
\label{fig:res-mass-varying}
\end{figure}

Lastly, the entropy-based method is applied to simulation data of eruptions with a range of SO\textsubscript{2} magnitudes to determine the impact's sensitivity with respect to the eruption strength. Figure~\ref{fig:res-mass-varying} shows the mean difference of TREFHT using eruption magnitudes of 10, 5 and 3 Tg.
In the 5 Tg case, small statistically significant cooling was found between September 26, 1992 and November 19, 1992 with an estimated temperature difference of $-0.391$K with a 99\% confidence interval between $-0.052$K and $-0.679$K. No impact was captured in the 3 Tg case.

\subsection{Pathway for temperate north surface cooling} \label{subsec:results-pathway}
The hypothesized pathway for temperate north surface cooling comes from an increase in aerosol optical depth in the region due to the Mt. Pinatubo eruption, followed by a decrease in solar flux at the surface. Full pathway DAGs are constructed by using AEROD\_v, FSDSC and TREFHT as variable constraints,  all zonal regions as spatial constraints, 4 years of temporal data, and a $t$-score of 1.0 to identify impact nodes. The impact pathway DAG is constructed by using the temperate north surface cooling impact identified in Section~\ref{subsec:results-temp} as the final impact node. The final impact node is used to constrain the graph by only adding nodes and edges that are upstream from the temperate north surface cooling between April and November of 1992. Table~\ref{tab:res-pathways-counts} shows the number of nodes and edges generated from a daily, monthly and entropy-based DAG.
\begin{table*}[htbp]
\centering
\caption{Full pathway DAGs are constructed using AEROD\_v, FSDSC and TREFHT as variable constraints, all zonal regions as spatial constraints, 4 years of temporal data, a $t$-score of 1.0 to identify impacts, and daily, monthly and entropy-based temporal means. The impact pathway DAG also uses the final impact node TREFHT from Section~\ref{subsec:results-temp} as a constraint. Entropy-based temporal means produce the least amount of nodes and edges.}
\label{tab:res-pathways-counts}
\begin{tabular}{p{4.0cm}p{4.0cm}p{2.0cm}p{2.0cm}}
\hline\noalign{\smallskip}
Pathway DAG Type & Temporal feature & Nodes & Edges \\
\noalign{\smallskip}\hline\noalign{\smallskip}
Full & Daily & 22303 & 127995 \\
 & Monthly & 789 & 3864 \\
 & Entropy-based & 288 & 937 \\
\noalign{\smallskip}\hline\noalign{\smallskip}
Impact & Daily & 8525 & 53408 \\
 & Monthly & 297 & 1546 \\
 & Entropy-based & 68 & 229 \\
\noalign{\smallskip}\hline\noalign{\smallskip}
\end{tabular}
\end{table*}
In this case, the entropy-based method produced the least number of nodes and edges. Also, the entropy-based nodes are directly linked to constant entropy between the eruption and counterfactual datasets meaning they are more likely to be correlated to impacts from the eruption dataset.

\begin{figure}[htbp]
\centering
\includegraphics[width=0.8\textwidth]{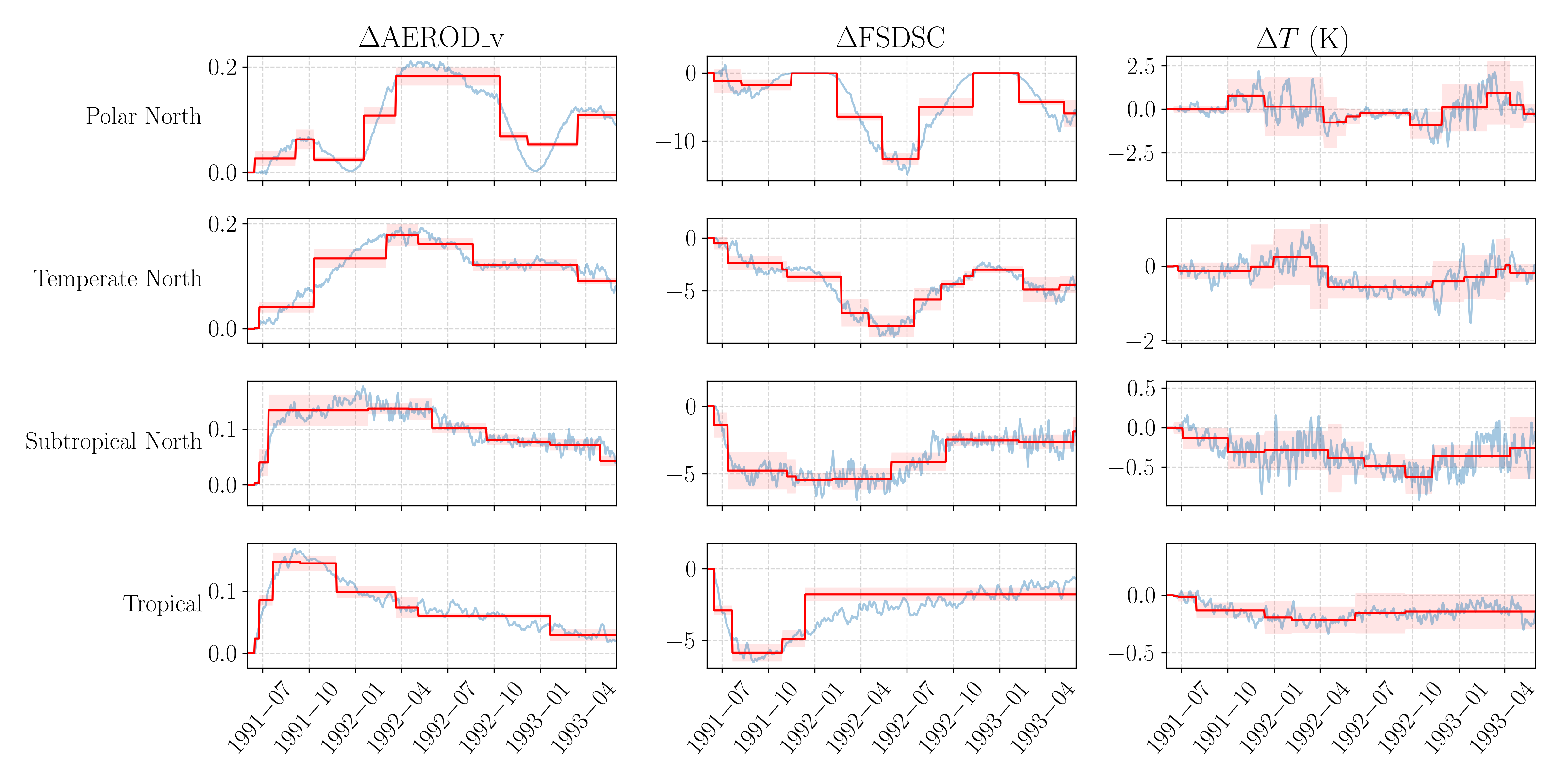}
\caption{The mean difference for AEROD\_v, FSDSC and TREFHT in four regions are computed in a 9-member ensemble with and without the Mt. Pinatubo eruption. The daily mean difference between the eruption and the counterfactual is in blue, the temporal mean in each entropy-based time interval is in red, and a 99\% confidence interval (CI) is in a red shade. The entropy-based method tracks shifts in irregularity between the two datasets. AEROD\_v and FSDSC have higher statistical significance when compared to TREFHT and impacts are much easier to identify.}
\label{fig:res-pathway-details}
\end{figure}

Figure~\ref{fig:res-pathway-details} shows the time series plots for all variables and relevant zonal regions and provides an overview of the features found by the entropy-based method.
The entropy-based method is largely tracking patterns present in both the eruption and counterfactual datasets. Also, AEROD\_v and FSDSC have higher statistical significance across the time series when compared to TREFHT. Impacts are much more difficult to identify in TREFHT.

\begin{figure}[htbp]
\centering
\includegraphics[width=0.8\textwidth]{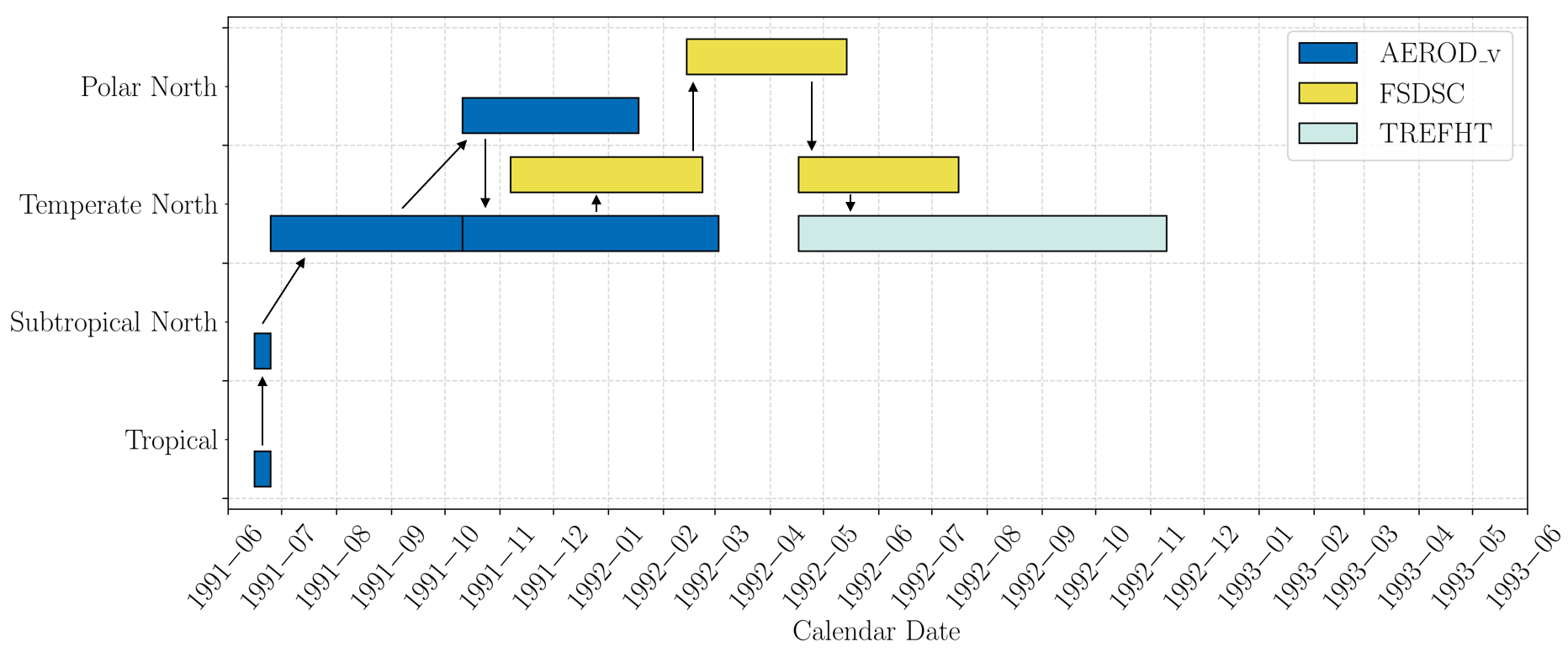}
\caption{The source-impact pathway DAG contains the path in the graph with the largest $t$-score starting from the final impact node, temperate north surface cooling (TREFHT), and ending at the source node, an increase in total aerosol optical depth in the tropics (AEROD\_v). The path travels rapidly into the northern regions and moves back and forth between temperate north and polar north before the final impact in 1992.}
\label{fig:res-pathway}
\end{figure}

The source-impact pathway DAG is constructed by using the first AEROD\_v impact node found in the full pathway DAG as the source node and traversing the path with the largest $t$-score starting from the final impact node. Figure~\ref{fig:res-pathway} shows the nodes and edges that were constructed in space-time while Table~\ref{tab:res-pathway} provides a list of the nodes along with their impact magnitude and $t$-score.
\begin{table*}[htbp]
\centering
\caption{Information about the nodes of the source-impact pathway DAG with the largest $t$-score are shown starting from the final impact node, temperate north surface cooling (TREFHT), and ending at the source node, an increase in total aerosol optical depth in the tropics (AEROD\_v). There is a large change in AEROD\_v in the northern regions which is associated with a decrease in solar flux at the surface (FSDSC) and a decrease in TREFHT in 1992.}
\label{tab:res-pathway}
\begin{tabular}{p{2.0cm}p{3.1cm}p{4.2cm}p{3.8cm}p{1.3cm}}
\hline\noalign{\smallskip}
Variable & Region & (Start, End) & Impact (99\% CI) & $t$-score \\
\noalign{\smallskip}\hline\noalign{\smallskip}
AEROD\_v & Tropical & (1991-06-16, 1991-06-25) & 0.024 (0.023,0.024) & 94.427 \\
AEROD\_v & Subtropical North & (1991-06-16, 1991-06-25) & 0.003 (0.001,0.004) & 6.249 \\
AEROD\_v & Temperate North & (1991-06-25, 1991-10-11) & 0.041 (0.031,0.051) & 13.629 \\
AEROD\_v & Polar North & (1991-10-11, 1992-01-18) & 0.024 (0.020,0.029) & 18.052 \\
AEROD\_v & Temperate North & (1991-10-11, 1992-03-03) & 0.134 (0.116,0.151) & 25.251 \\
FSDSC & Temperate North & (1991-11-07, 1992-02-23) & -3.665 (-4.140,-3.189) & -25.842 \\
FSDSC & Polar North & (1992-02-14, 1992-05-14) & -6.392 (-6.933,-5.851) & -39.644 \\
FSDSC & Temperate North & (1992-04-17, 1992-07-16) & -8.374 (-9.401,-7.347) & -27.367 \\
TREFHT & Temperate North & (1992-04-17, 1992-11-10) & -0.560 (-0.864,-0.257) & -6.190 \\
\noalign{\smallskip}\hline\noalign{\smallskip}
\end{tabular}
\end{table*}
The pathway focuses on the large change in AEROD\_v which travels up into the northern temperate and polar regions around the winter of 1991 and reduces the solar flux at the surface during the winter, spring and summer of 1992. This is then connected with the temperate north surface cooling discussed in Section~\ref{subsec:results-temp}. Statistically significant impacts are identified very early in the northern regions. Polar impacts with higher statistical significance when compared to temperate north impacts are also found which causes the path to move back and forth between temperate north and polar north before the final impact in 1992. This also suggests a significant amount of aerosol mixing in the northern regions before 1992. This is not to suggest that pathways must travel through the polar north to reach the final impact node. The impact pathway DAG contains more pathways with more nodes that connect the impacts shown in Figure~\ref{fig:res-pathway-details}.

\section{Conclusions} \label{sec:conc}
In this paper, a new entropy-based feature selection method is developed for capturing impacts and constructing data-driven dependency graph models in Earth system model datasets with and without external forcing. The goal is to reduce the complexity of the dataset and only focus on the most important features related to a regional impact of interest. The method is tested on a Mt. Pinatubo eruption simulation ensemble where it was shown to identify and quantify temperate north surface cooling and connect it to a decrease in solar flux at the surface and an increase in aerosol optical depth in the tropics. It was also shown that the impact remained statistically significant at 5 Tg as the SO\textsubscript{2} magnitude was decreased but no impact was captured at 3 Tg. Though the methods were constructed for Earth system modeling data, they are extensible to any nonlinear dynamical system with high frequency spatio-temporal data.

The temporal feature selection method uses cross-fuzzy entropy and changepoint detection to capture time intervals of constant entropy. Many of the time intervals were found to be robust with increasing ensemble size but it was not clear whether shifts in similarity of patterns were caused by the eruption or seasonal variability. Nevertheless, it was shown that grouping time intervals based on entropy reduces the likelihood of identifying impacts that are not actually significant, or false positives. It may be possible to construct a multi-scale, adaptive approach which can work on instantaneous data and isolate features more accurately. Since the method works on a sliding window of data, it is also possible to implement this in situ within an Earth system model for adaptive impact monitoring and control. Spatial feature selection without the loss of regional labels also remains to be explored.

The entropy-based impact pathway DAG that was constructed has two orders of magnitude less nodes than the daily impact pathway DAG and produced a direct link between impact in aerosol optical depth in the tropics and temperate north surface cooling by identifying statistically significant impacts. It remains to be seen whether the reduced dataset is sufficient to produce or tune a data-driven model for causal inference or accurate predictions of regional impacts subject to a similar forcing. However, the graph alone provides a unique insight into how impacts may propagate in space-time and could be used in future studies of Mt. Pinatubo.

\section*{Data and code availability}
The data that support the findings of this study are available by contacting the corresponding author. Data from the full E3SMv2-SPA simulation campaign including pre-industrial control, historical, and Mt. Pinatubo ensembles will be hosted at Sandia National Laboratories with location and download instructions announced on \url{https://www.sandia.gov/cldera/e3sm-simulations-data} when available.

\section*{Acknowledgments}
This work was supported by the Laboratory Directed Research and Development program at Sandia National Laboratories, a multimission laboratory managed and operated by National Technology and Engineering Solutions of Sandia, LLC, a wholly  owned subsidiary of Honeywell International, Inc., for the U.S. Department of Energy’s National Nuclear Security Administration under contract DE-NA-0003525. This research used resources of the National Energy Research Scientific Computing Center (NERSC), a U.S. Department of Energy Office of Science User Facility using NERSC award BER-ERCAP0026535.

The authors thank Hunter Brown, Thomas Ehrmann, Benjamin Hillman, Kara Peterson, and Benjamin Wagman for designing the prognostic aerosol modifications and model initialization needed to provide a realistic Mt. Pinatubo eruption ensemble with limited intra-ensemble variability.

\section*{Disclaimer}
This paper describes objective technical results and analysis. Any subjective views or opinions that might be expressed in the paper do not necessarily represent the views of the  U.S. Department of Energy or the United States Government.

\bibliographystyle{plain}
\bibliography{references}

\end{document}